\documentclass[aip,superscriptaddress,reprint,jap]{revtex4-2}
\pdfoutput=1
\usepackage{xfrac}
\usepackage{svg}
\usepackage{tikz}
\usepackage{amssymb}
\usepackage{psfrag}
\usepackage{textcomp}
\usepackage{amsmath}
\usepackage{amsfonts}
\usepackage{graphicx,bm,epstopdf}
\usepackage[caption=false]{subfig}
\usepackage{comment}
\usepackage{verbatim}
\usepackage{gensymb}
\usepackage{floatrow}
\usepackage{float}
\usepackage{xcolor}
\usepackage[version=4]{mhchem}
\usepackage{upgreek}
\usepackage{lineno}
\usepackage{threeparttable}
\usepackage[colorlinks]{hyperref}

\hypersetup{
    colorlinks=true,
    citecolor=blue,
    linkcolor=blue,
    filecolor=magenta,      
    urlcolor=blue,
    }

\begin{document}
\title{Non-reciprocal Magnetoresistances in Chiral Tellurium}
\author{Shuchen Li}
\affiliation{%
Department of Materials Science and Engineering and Materials Research Laboratory, The Grainger College of Engineering, University of Illinois Urbana-Champaign, Urbana, Illinois 61801, USA
}%

\author{Chang Niu}
\affiliation{%
Birck Nanotechnology Center, Purdue University, West Lafayette, IN 47907, United States.
}%
\affiliation{%
Elmore Family School of Electrical and Computer Engineering, Purdue University, West Lafayette, IN 47907, United States.
}%

\author{Peide D. Ye}
\affiliation{%
Birck Nanotechnology Center, Purdue University, West Lafayette, IN 47907, United States.
}%
\affiliation{%
Elmore Family School of Electrical and Computer Engineering, Purdue University, West Lafayette, IN 47907, United States.
}%
\author{Axel Hoffmann}
    \email{axelh@illinois.edu}
\affiliation{%
Department of Materials Science and Engineering and Materials Research Laboratory, University of Illinois Urbana-Champaign, Urbana, Illinois 61801, USA
}%

\begin{abstract}
Materials with broken fundamental symmetries, such as chiral crystals, provide a rich playground for exploring unconventional spin-dependent transport phenomena. The interplay between a material's chirality, strong spin-orbit coupling, and charge currents can lead to complex non-reciprocal effects, where electrical resistance depends on the direction of current and magnetic fields. In this study, we systematically investigate the angular dependencies of magnetoresistance in single-crystalline chiral Tellurium (Te). We observe distinct non-reciprocal magnetoresistances for magnetic fields applied along three orthogonal directions: parallel to the current along the chiral axis (\textbf{z}), in the sample plane but perpendicular to the current (\textbf{y}), and out of the sample plane (\textbf{x}). Through detailed analysis of the chirality- and thickness-dependence of the signals, we successfully disentangle multiple coexisting mechanisms. We conclude that the Edelstein effect, arising from the chiral structure's radial spin texture, is responsible for the non-reciprocity along the \textbf{z}-axis. In contrast, the chirality-independent signal along the \textbf{y}-axis is attributed to the Nernst effect, and the non-reciprocity along the \textbf{x}-axis may originate from the orbital magnetization. These findings elucidate the complex interplay of spin, orbital, and thermal effects in Te, providing a complete picture of its non-reciprocal transport properties.
\end{abstract}

\maketitle

\section{Introduction}
The breaking of fundamental symmetries in crystalline solids is a cornerstone for the discovery and manipulation of novel quantum phenomena\cite{RevModPhys.76.323}. In materials that lack a center of inversion, the interplay between the spin of an electron and its momentum, governed by strong spin-orbit coupling (SOC), gives rise to a host of unconventional transport effects that are forbidden in centrosymmetric systems\cite{Manchon2015, PhysRevLett.110.067207, 10716676,PhysRevApplied.22.044043, PhysRevB.110.024426}. One of the most fascinating of these is non-reciprocal charge transport, where the electrical resistance is not the same when the current direction is reversed. This effect serves as a powerful probe of symmetry-breaking and has significant potential for applications in next-generation electronics\cite{He2018, Tokura2018, Suárez-Rodríguez2025}.

A primary mechanism for generating non-reciprocal transport is through the creation of a net spin polarization in the material. In systems with strong Rashba-type SOC, such as the polar semiconductor BiTeBr, an applied magnetic field can asymmetrically shift the spin-split Fermi surfaces, leading to a current-direction-dependent resistance, often termed unidirectional magnetoresistance (UMR) or bilinear magnetoelectric resistance\cite{Ideue2017}. This effect can be intuitively understood as the interplay between the external magnetic field and current-induced net spin polarizations that arise from the out-of-equilibrium shift of the spin-momentum-locked Fermi contours\cite{PhysRevLett.124.027201, He2018}. The magnitude of this non-reciprocity is directly tied to the strength of the underlying spin-orbit interaction, making it a valuable tool for quantifying spin-splitting. Moreover, its detailed angular dependence makes UMR an exceptionally sensitive probe of the complete three-dimensional spin texture, capable of revealing complex features beyond the standard Rashba model, such as the momentum-dependent out-of-plane spin components observed in SrTiO$_3$ based 2 dimensional electron gas\cite{PhysRevLett.120.266802}. Distinct from this spin-based mechanism, non-reciprocity can also arise from orbital degrees of freedom. In symmetry-breaking systems such as twisted bilayer graphene\cite{He2020}, a net magnetization in the out-of-plane direction were observed due to a current-induced imbalance in the distribution of orbital magnetic moments. In chiral crystals, the structural helicity allows for the generation of a current-induced orbital magnetization\cite{hua2025interplayorbitalspinmagnetization}. While both effects are driven by the applied current and share the same phenomenological scaling ($V \propto I \cdot B$), they stem from fundamentally different angular momentum contributions—spin versus orbital—providing a unique opportunity to disentangle these quantum degrees of freedom through their distinct angular dependencies.

Tellurium (Te), an elemental material that has a chiral crystal structure with three tellurium atoms forming a spiral-shaped covalently bonded atomic chain in each unit cell, stands out as a compelling material for exploring non-reciprocal transport through the interplay among structural chirality and strong SOC\cite{PhysRevResearch.3.023111, PhysRevLett.124.136404, PhysRevB.101.205414}. The helicity of this atomic chain imparts a well-defined chirality, breaking both inversion and mirror symmetries. This unique crystallographic property, combined with a strong intrinsic SOC, lifts spin degeneracy and leads to a complex band structure with a radial spin texture at the Fermi surface. When a charge current is applied along the chiral axis, the Edelstein effect is predicted to induce a net spin polarization collinear with the current, with opposite polarities for right- and left-handed Te\cite{EDELSTEIN1990233} as shown in Fig.~\ref{Fig1} (a), also making Te an ideal platform for investigating novel spintronic phenomena. This is a critical goal for developing next-generation, low-power electronic devices\cite{Calavalle2022, 9427163, doi:10.1021/acs.nanolett.3c00780}. Moreover, with an application of an external  magnetic field, non-reciprocal magnetoresistance have been found and reported in Te, while a comprehensive understanding of how structural chirality governs magnetotransport responses under various magnetic field orientations is still developing\cite{PhysRevResearch.3.023111, Calavalle2022, doi:10.1021/acs.nanolett.3c00780, PhysRevB.111.024405, Ma2024}. In this paper, we systematically investigate the non-reciprocal magnetoresistance in single-crystalline Te. We identify and disentangle three distinct non-reciprocal effects by analyzing their unique dependencies on crystal chirality, sample thickness, and magnetic field direction, attributing their origins to the Edelstein effect, the Nernst effect\cite{PhysRevLett.108.106602,PhysRevLett.110.067207}, and the orbital magnetization\cite{hua2025interplayorbitalspinmagnetization, 7nxc-j62y}, respectively.

\section{Measurement}
\begin{figure}[h!tbp] 
    \centering
    \includegraphics[width=\textwidth]{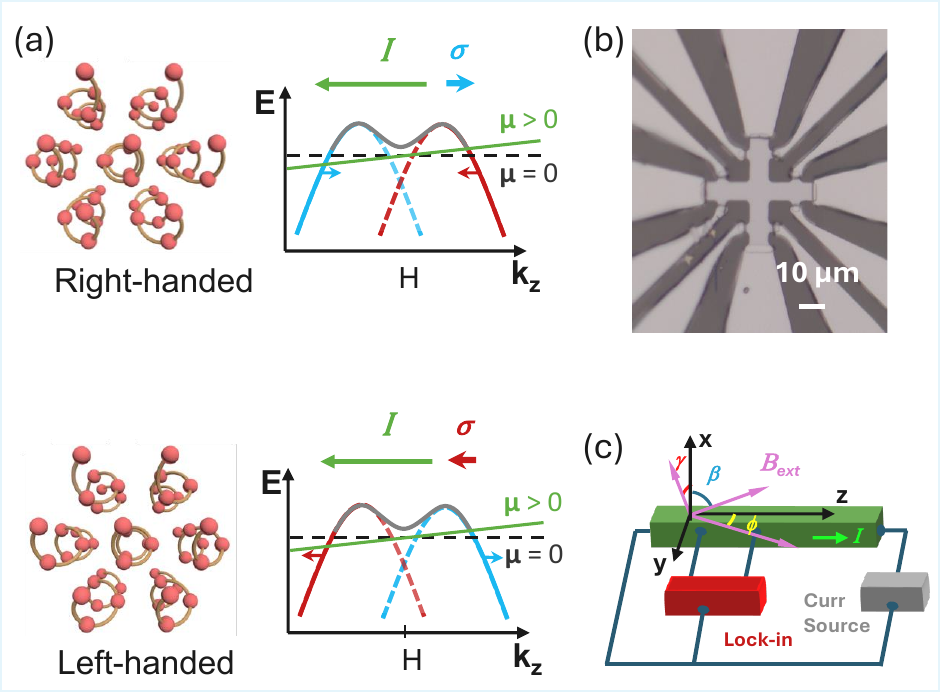}
    \caption{(a)Diagrams showing left-handed and right-handed Te crystal structures and the corresponding spin polarizations due to Edelstein effect.(b)Microscopic image of a fabricated Te device, with current channels both along and perpendicular to the chiral axis. (c)Schematics showing the measurement setup under various field scans. }
    \label{Fig1}
\end{figure}

Single crystalline Te flakes were grown by using the hydrothermal method\cite{doi:10.1021/acs.nanolett.3c00780}, and we used lithography and deposited Pt electrodes for transport measurements, in which a current is applied along +\textbf{z}, and \textbf{y}(\textbf{x}) direction is orthogonal to the current-direction in (out) of the device plane[Fig.~\ref{Fig1} (b)]. At the same time, a magnetic field $B$ is applied in the \textbf{zy}(\textbf{xz}, \textbf{xy}) plane, and we detected the longitudinal resistance using a lock-in amplifier at various angles $\phi$($\beta$, $\gamma$), where $\phi$($\beta$, $\gamma$) is the angle between $B$ and +\textbf{z}(+\textbf{x},+\textbf{x}). In this paper, $\phi$($\beta$, $\gamma$) = 0$^\circ$ when $B$ is parallel to +\textbf{z}(+\textbf{x},+\textbf{x}) and $\phi$($\beta$, $\gamma$) = 90$^\circ$ when $B$ along +\textbf{y}(+\textbf{z},+\textbf{y}) axis, as shown in Fig.~\ref{Fig1} (c).

\begin{figure}[h!tbp] 
    \centering
    \includegraphics[width=\textwidth]{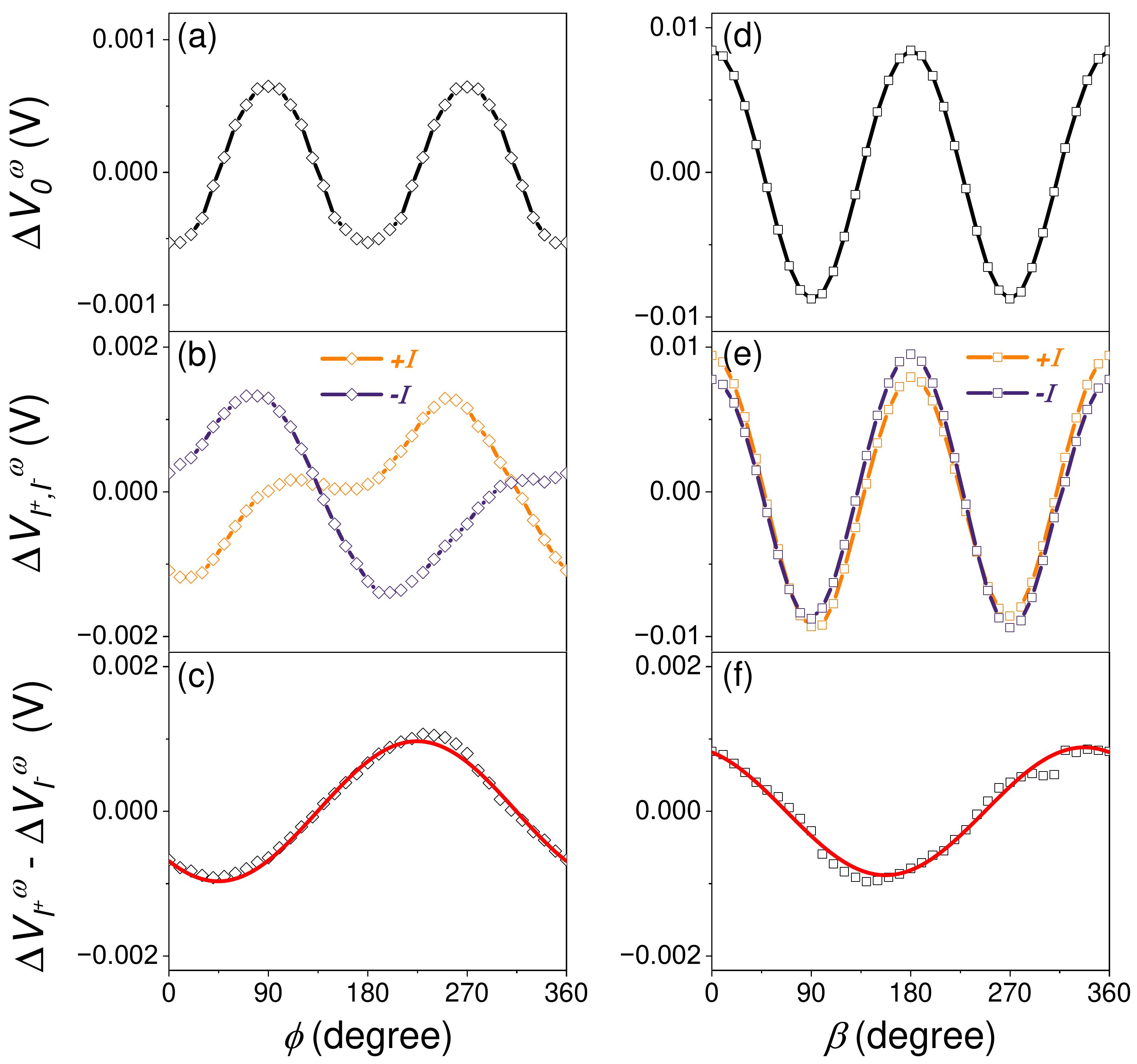}
    \caption{(a) and (d)$\beta$ and $\phi$ scans of mangetoresistance of Te with pure ac currents. (b) and (e)$\phi$, $\beta$ scans of Te with positive(orange) and negative(purple) dc offsets at 5 $K$. (c) and (f)Angular dependencies of $\Delta V_{I^+}^\omega$ - $\Delta V_{I^-}^\omega$ to extract components with sine and cosine angular dependencies.}
    \label{Fig2}
\end{figure}

We started our measurement by applying an ac current $I_0 sin(\omega t)$ with $I_0 = 3e^{-4}$ $A$ along the chiral axis of Te (R1) and collected the longitudinal first harmonic signal $V_0^\omega$ as we rotated the magnetic field both in the $\phi$ and $\beta$ plane at 5 $K$. As shown in Fig.~\ref{Fig2} (a) and (d), we plot $\Delta V_0^\omega = V_0^\omega - V^\omega_{0,average}$ as a function of $\phi$ and $\beta$ at $B = 2$ $T$, and we found Te has largest magnetoresistance when $B$ is out-of-plane along \textbf{x} and smallest when $B$ is parallel with the chiral axis and the current direction. 

We then applied a positive dc offset $I_{dc}$ to the ac current, with $I = I_0 sin(\omega t) + I_{dc}$, and $I_0 = |I_{dc}| = 3e^{-4}$~$A$, and performed similar $\phi$ and $\beta$ scans. Interestingly, as shown in Fig.~\ref{Fig2} (b) and (e), we found the angular dependencies(orange curves) of $\Delta V_{I^+}^\omega = V_{I^+}^\omega - V_{I^+,average}^\omega$ were drastically changed with the application of a positive dc offset, and Te magnetoresistances were no longer even with the field, with clear non-reciprocal features appearing in both $\phi$ and $\beta$ scans. Specifically, in $\phi$ scan, we found $V_{I^+}^\omega$ is smaller when the $B$ is parallel to the current direction ($\phi$ = 0$^\circ$) and larger when they are antiparallel ($\phi$ = 180$^\circ$). Surprisingly, we have also observed a difference in $\Delta V_{I^+}^\omega$ when $\phi$ = 90$^\circ$ and 270$^\circ$, with a smaller $\Delta V_{I^+}^\omega$ at $\phi$ = 90$^\circ$. The two non-reciprocal behaviors of magnetoresistance in $\phi$ scan also appear in $\beta$ scans at $\beta$ =  90$^\circ$ and 270$^\circ$. Remarkably, we also observed another non-reciprocal behavior when the field is along \textbf{x} out-of-plane, as $\Delta V_{I^+}^\omega$ at $\beta$ = 0$^\circ$ and 180$^\circ$ is different. We then reversed the current direction by applying a negative $I_{dc}$ and observed behaviors of $\Delta V_{I^-}^\omega$ also reversed(purple curves in Fig.~\ref{Fig2} (d) and (e)) as expected. 

Based on our experimental observations and the general $D_3$ point group symmetry constraints, we generalized the electrical resistance of Te as

\begin{widetext}
    \begin{equation}
   R (I,B) = R_0[1 + B^2(\delta_z \cos\phi \sin(\theta) + \delta_y \sin\theta \sin\phi + \delta_x \cos\theta)^2 + \rho I^2] \\+ IB(\alpha_z \sin\theta \cos\phi + \alpha_y\sin\theta \sin\phi + \alpha_x\cos\phi)
   \label{R}
    \end{equation}
\end{widetext}

where $R_0$, $I$, and $B$ represent the resistance of Te at zero field, the current, and the external magnetic field. The coefficients $\delta_z$, $\delta_y$, and $\delta_x$ correspond to the normal magnetoresistances of Te when $B$ is along \textbf{z}, \textbf{y}, and \textbf{x} that will cause $R$ to either decrease or increase with the field, and are independent of current. On the other hand, $\alpha_z$, $\alpha_y$, and $\alpha_x$ correspond to the sizes of the non-reciprocal effects in Te for $B$ along \textbf{z}, \textbf{y}, and \textbf{x}, which are dependent on both $B$ and $I$, rectifying Te magnetoresistances. $\rho$ is the coefficient for thermal effect.

\begin{figure*}[h!tbp] 
    \centering
    \includegraphics[width=\columnwidth]{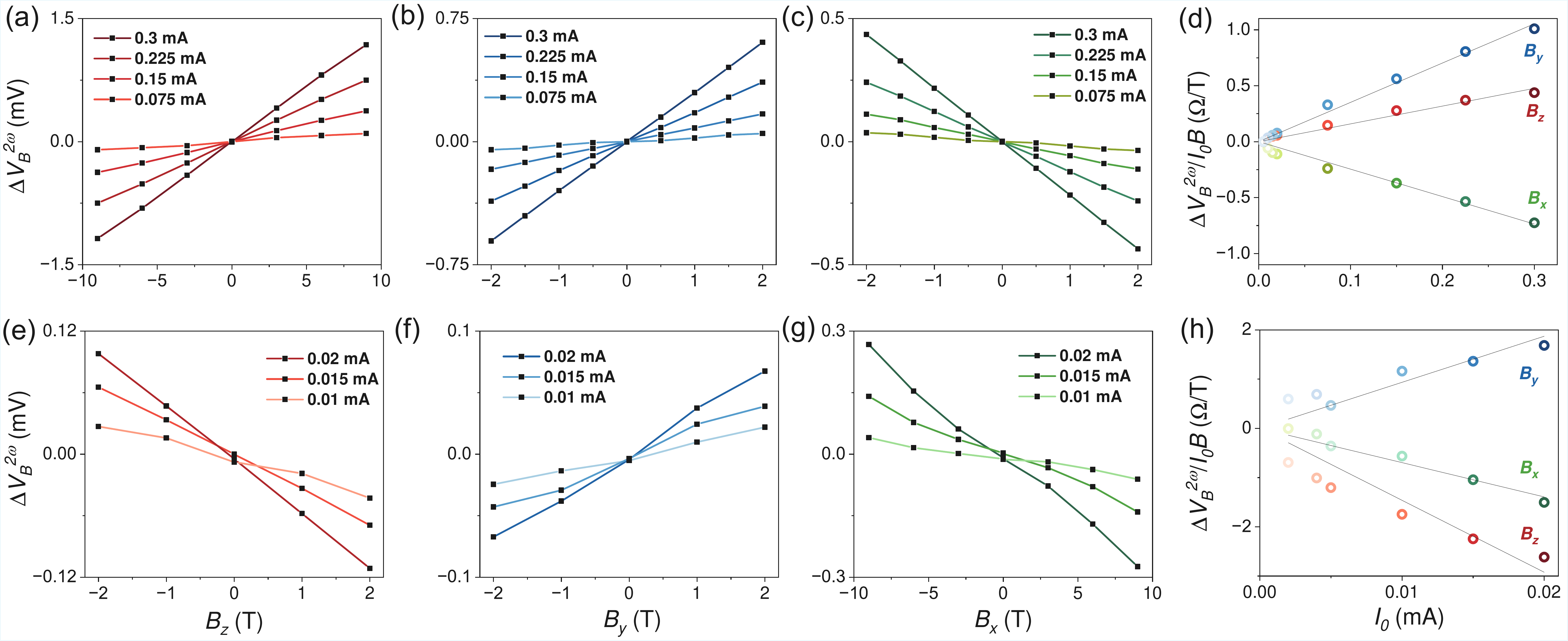}
    \caption{(a)-(c)$\Delta V_B^{2\omega}$ under $B_z$, $B_y$, and$B_x$ field sweep for Te(R1) at 5 $K$. (e)-(g)$\Delta V_B^{2\omega}$ under $B_z$, $B_y$, and$B_x$ field sweep for Te(L1) at 5 $K$. (d) and (h) $\Delta V_B^{2\omega}/(I_0 B)$ vs current for Te(R1) and (L1). Red, blue, and green dots correspond to $\Delta V_B^{2\omega}/(I_0 B)$ at fields along $B_z$, $B_y$, and $B_x$ at different current densities. }
    \label{Fig3}
\end{figure*}

Thus, when we applied a dc offset with $I = I_0 sin(\omega t) + I_{dc}$, the resultant longitudinal first harmonic signal $V^\omega$ would also be modified (see supplemental materials), giving rise to the observed non-reciprocal effects. To exclude other contributions in Te, we calculated and plotted $\Delta V_{I^+}^\omega$ - $\Delta V_{I^-}^\omega$ under the same field but opposite dc offsets in Fig.~\ref{Fig2} (c) and (f), and extracted $\alpha_z$ = -3.1 $\pm$ 0.2 k$\Omega/(A\cdot T)$, $\alpha_y$ = -3.6 $\pm$ 0.2 k$\Omega/(A\cdot T)$, and $\alpha_x$ = 3.8 $\pm$ 0.2 k$\Omega/(A\cdot T)$ by using Eq.1 from cosine and sine components in the $\phi$ and $\beta$ dependencies of $\Delta V^\omega$ easily.

 

\begin{figure}[htbp] 
    \centering
    \includegraphics[width=\textwidth]{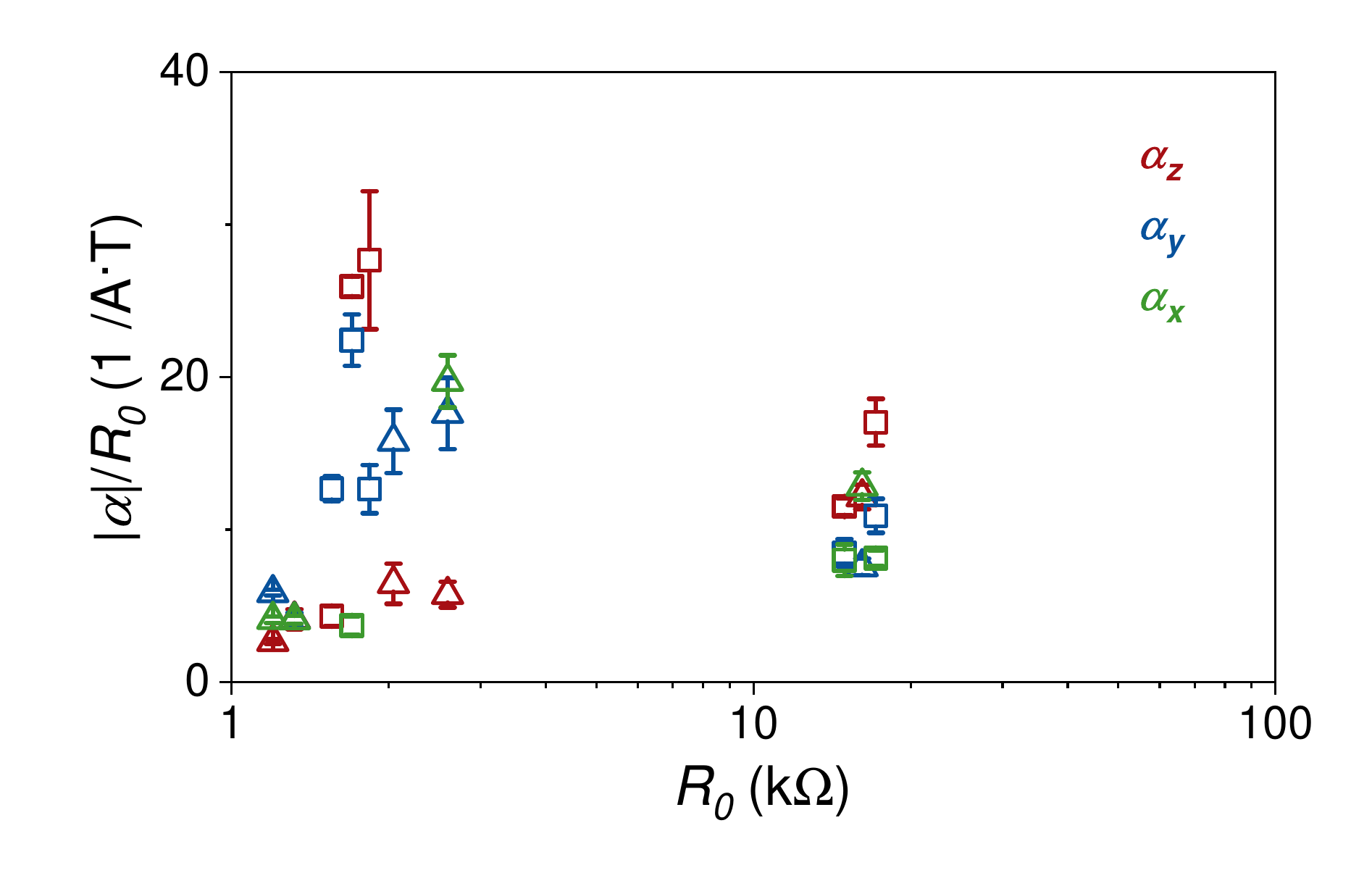}
    \caption{Ratio of absolute values of extracted $\alpha_z$, $\alpha_y$, and $\alpha_x$ and the corresponding device resistance $R_0$ as a function of $R_0$. Sqaures and triangles represent left and right handedness.}
    \label{Fig4}
\end{figure}

It is instructive to study also the non-linear second harmonic signal $V^{2\omega}$ proportional to the current, which also reflects the non-reciprocal effects and is independent of $I_{dc}$, and the analysis of it can be complementary to that of $V^{\omega}$. Fig.~\ref{Fig3} (a)-(c) showed $\Delta V_B^{2\omega}$ of Te (R1) as a function of $B_z$, $B_y$, and $B_x$ up to 9 T and 2 T due to system geometry restrictions, with various ac amplitude $I_0$ at 5 $K$, where $\Delta V_B^{2\omega} = (V^{2\omega}(\pm B) - V^{2\omega}(\mp B))/2$ is the absolute change in second harmonic signal. Clear linear relationships with all $B_z$, $B_y$, and $B_x$ can be observed from all $I_0$, agreed well with Eq.~\ref{R}, confirming three non-reciprocal behaviors observed in first harmonic magnetoresistance measurements. Fig.~\ref{Fig3} (d) shows $\Delta V_B^{2\omega}/(I_0 B)$ as a function of $I_0$ for $B_z$(red), $B_y$(blue), and $B_x$(green), and the slopes are directly proportional to the sizes of $\alpha_z$, $\alpha_y$, and $\alpha_x$. We calculated $\alpha_z$ = -3.2 $\pm$ 0.1 k$\Omega/(A\cdot T)$, $\alpha_y$ = -7 $\pm$ 0.1 k$\Omega/(A\cdot T)$, and $\alpha_x$ = 4.9 $\pm$ 0.1 k$\Omega/(A\cdot T)$, comparable to the results derived from the magnetoresistance measurements.
Similarly, we measured $V^{2\omega}$ for Te (L1) as a function of $I_0$ and fields. Fig.~\ref{Fig3} (e)-(g) showed $\Delta V_B^{2\omega}$ as a function of $B_z$, $B_y$, and $B_x$ for different current densities along the chiral axis. Clearly, the slope of the $B_z$-dependency of $\Delta V_B^{2\omega}$ changed sign when Te chirality changed. However, the slope of $B_y$ and $B_x$ dependencies stayed the same. We then plot $\Delta V_B^{2\omega}/(I_0 B)$ as a function of $I_0$ in Fig.~\ref{Fig3} (h) and extract $\alpha_z$, $\alpha_y$, and $\alpha_x$ to be 292.3 $\pm$ 26.4 k$\Omega/(A\cdot T)$, -187.1 $\pm$ 19.2 k$\Omega/(A\cdot T)$, and 139.3 $\pm$ 8.8 k$\Omega/(A\cdot T)$. We noticed a significant difference in the magnitudes of all coefficients between the two samples, which is likely related to their different resistances (1200~$\ohm$ vs 17200 $\ohm$), so we studied the thickness dependence of $\alpha_z$, $\alpha_y$, and $\alpha_x$ by performing the second harmonic measurements across a wide range of Te samples with various thicknesses. Table~\ref{table_coefficients} recorded the calculated coefficients together with the resistances of corresponding devices, which is directly related to the sample thicknesses. First, we found $\alpha_z$ indicates a clear chirality-dependent origin with both positive and negative signs among the samples, while $\alpha_y$ is always negative in all the samples measured, suggesting a chirality-independent origin. We also noticed that the sign of $\alpha_x$ also varied across different Te samples, but did not tie to that of $\alpha_z$, which means $\alpha_x$ could have opposite signs even if $\alpha_z$ had the same sign, for example from device 1 and 2, suggesting distinct physical origins for $\alpha_z$ and $\alpha_x$. We then plotted the ratio between absolute values of extracted $\alpha_z$, $\alpha_y$, and $\alpha_x$ and the corresponding device resistance $R_0$ as a function of $R_0$ in Fig.~\ref{Fig4}, which are similar for all three coefficients across different resistances,indicating intrinsic bulk origins behind.

\begin{table}[h!]
    \centering
    \begin{ruledtabular}
    \begin{tabular}{ccccc}
        Device & \textbf{$R_0$}  [k$\ohm$] & \textbf{$\alpha_z$} [k$\Omega/(\mathrm{A}\cdot \mathrm{T})$] & \textbf{$\alpha_y$} [k$\Omega/(\mathrm{A}\cdot \mathrm{T})$] & \textbf{$\alpha_x$} [k$\Omega/(\mathrm{A}\cdot \mathrm{T})$] \\
        \hline
        L1 & 17.2 & 292.3 $\pm$ 26.4 & -187.1 $\pm$ 19.2 & 139.3 $\pm$ 8.8 \\
        L2 & 14.9 & 172 $\pm$ 8.1 & -126.5 $\pm$ 13.5 & -119.2 $\pm$ 15.5\\
        R1 & 1.2 & -3.2 $\pm$ 0.1 & -7 $\pm$ 0.1 & 4.9 $\pm$ 0.1 \\
        R2 & 1.3 & -6.6 $\pm$ 0.9 & -9.2 $\pm$ 0.2 & -8.5 $\pm$ 0.6 \\
        L3 & 1.8 & 50.8 $\pm$ 8.3  & -23.3 $\pm$ 3.0 &  \\
        L4 & 1.7 & 44.1 $\pm$ 1.1 & -38.1 $\pm$ 3.0 & -6.3 $\pm$ 1.1 \\
        R3 & 2.0 & -13.2 $\pm$ 2.7 & -32 $\pm$ 4.2 & \\
        L5 & 1.6 & 6.7 $\pm$ 1.0 & -19.7 $\pm$ 1.3 & \\
        R4 & 2.6 & -14.9 $\pm$ 2.2 & -45.7 $\pm$ 6.1 & -51.1 $\pm$ 4.4\\ 
        R5 & 16.2 & -196 $\pm$ 12.9 & -122 $\pm$ 8.8 & 207.5 $\pm$ 14.5 \\
    \end{tabular}
    \end{ruledtabular}
    \caption{Resistances $R_0$ of different Te devices and extracted non-reciprocal effect coefficients $\alpha_z$, $\alpha_y$, $\alpha_x$ from second harmonic voltages. Right or left chirality is indicated by R and L with the device number followed }
    \label{table_coefficients}
\end{table}

\section{Discussion}
We would like to dicuss possible origins behind the non-reciprocal magnetoresistances observed in Te, for fields along \textbf{z}, \textbf{y} and \textbf{x} when current is along the chiral axis.

First, the non-reciprocal magnetoresistance along the chiral axis displayed a clear chirality-dependent behavior which changed signs as we changed Te handedness, and we attributed that to the chirality-induced spin polarizations along the current directions due to the Edelstein effect and Te radial spin textures~\cite{Calavalle2022}. Also, magnitude of $\alpha_z$ decreased as Te thicknesses increased, further confirming the bulk effect from the radial spin textures.

However, the non-reciprocal magnetoresistances when $B$ is along y that always had the same polarities regardless of the handedness, and the corresponding coefficient $\alpha_y$ is always negative for all the samples measured, strongly suggesting a chirality-independent origin behind, and we proposed Nernst effect could explain the observations, that is a thermal gradient $\Delta T$ generated out-of-plane could give rise to different magnetoresistances for $\pm B_y$, whose magnitude is proportional to $\Delta T \times B_y$. We note that thermally generated $\Delta T$ does not reverse when we reversed the DC bias, and $\alpha_y$ would always be negative.

Finally, we attribute the non-reciprocal magnetoresistance observed under an out-of-plane magnetic field ($B \parallel x$) to the current-induced orbital magnetization. Unlike an ideal solenoid, the distorted helical crystal structure of Te generates not only a longitudinal moment ($M_z$) along the chiral axis\cite{PhysRevResearch.3.023111} but also a significant transverse orbital magnetization ($M_y$) perpendicular to the chiral axis\cite{hua2025interplayorbitalspinmagnetization}. Consequently, the interaction between this transverse moment $M_y$ and the out-of-plane field ($B_x$) induces a rectification voltage $V_{orb,y} \propto M_y \times B_x$. Crucially, because $M_y$ is fixed to the crystal lattice, its sign in the laboratory frame depends on the specific azimuthal orientation of the flake. We attribute the observed variability in the sign of $\alpha_x$ among samples of the same chirality to the random nature of device fabrication, where flakes may be deposited either 'face-up' or 'face-down' (corresponding to a $180^{\circ}$ rotation about the $z$-axis). Such a rotation reverses the transverse component $M_y$ relative to the external field, flipping the sign of the measured non-reciprocity, whereas the chirality-dependent longitudinal response remains unaffected.



\section{Conclusion}
In summary, we have systematically characterized the angular-dependent magnetoresistance in single-crystalline chiral Te flakes. Our measurements reveal three distinct non-reciprocal transport phenomena when a current is applied along the chiral axis. We have successfully disentangled their physical origins based on their unique dependencies on crystal chirality and magnetic field orientation. The chirality-dependent non-reciprocity along the transport direction is attributed to the Edelstein effect acting on the radial spin texture of Te's surface states. A second, chirality-independent effect is explained by the Nernst effect arising from a thermally generated gradient. Finally, a third non-reciprocity, observed with an out-of-plane field, is consistent with the influence of an intrinsic orbital magnetization. The pronounced thickness dependence of both the Edelstein effect and the orbital magnetization contributions underscores their origin at the bulk of the material. These findings not only provide a comprehensive picture of the complex interplay between charge, spin, and orbital degrees of freedom in a chiral conductor but also highlight the potential of tellurium for developing novel spintronic and orbitronic devices.

\section{Acknowledgements}
The authors would like to thank Paul Rutherford, Mandela Mehraeen, and Shulei Zhang (Case Western Reserve University) for insightful discussions regarding the origins behind Te's non-reciprocal magnetoresistance behaviors, and Ruihao Liu (Tsinghua University) for initiating the discussion between the authors and fostering this collaborative work. Support for the magnetoresistance measurements, data analysis, and manuscript preparation was provided by the Air Force Office of Scientific Research (AFOSR) MURI program under award number FA9550-23-1-0311.

\bibliography{Te}

\end{document}